\documentclass[12pt,preprint]{aastex}

\shorttitle{Do we need to know the temperature in prestellar cores?}
\shortauthors{Pavlyuchenkov et al.}

\begin{document}
\title{Do we need to know the temperature in prestellar cores?}

\author{Ya. Pavlyuchenkov}
\affil{Max Planck Institute for Astronomy,
K\"onigstuhl 17, D-69117 Heidelberg, Germany}
\email{pavyar@mpia.de}

\author{Th. Henning}
\affil{Max Planck Institute for Astronomy,
K\"onigstuhl 17, D-69117 Heidelberg, Germany}
\email{henning@mpia.de}

\and

\author{D. Wiebe}
\affil{Institute of Astronomy of the Russian Academy of
Sciences,
48, Pyatnitskaya Str., Moscow, 119017, Russia}
\email{dwiebe@inasan.ru}

\begin{abstract}
Molecular line observations of starless (prestellar) cores combined with a chemical evolution modeling and radiative transfer calculations are a powerful tool to study the earliest stages of star formation.  However, conclusions drawn from such a modeling may noticeably depend on the assumed thermal structure of the cores. The assumption of isothermality, which may work well in chemo-dynamical studies, becomes a critical factor in molecular line formation simulations. We argue that even small temperature variations, which are likely to exist in starless cores, can have a non-negligible effect on the interpretation of molecular line data and derived core properties. In particular, ``chemically pristine'' isothermal cores (low depletion) can have centrally peaked C$^{18}$O and C$^{34}$S radial intensity profiles, while having ring-like intensity distributions in models with a colder center and/or warmer envelope assuming {\it the same\/} underlying chemical structure. Therefore, derived molecular abundances based on oversimplified thermal models may lead to a mis-interpretation of the line data.
\end{abstract}

\keywords{astrochemistry --- line: profiles --- stars: formation --- ISM: molecules}

\section{Introduction}

The cooling in starless (or prestellar) cores is believed to be very effective, so it is customary to model these cores as isothermal at the temperature of around 9--10\,K. In general, this assumption is confirmed both observationally and theoretically. However, different authors disagree on possible minor deviations from the uniform temperature, which may exist in these objects.

More precisely, various attempts to model the gas thermal balance in starless cores basically reach the same conclusion that the gas and dust temperatures are well coupled in the core center and have values of 5--7\,K \citep[e.~g.][]{goldsmith,zucconi,galli,lesaffre}. At a low density periphery, heating processes due to cosmic rays and some UV irradiation take over, and the gas temperature rises to 15\,K and higher (depending on the UV field strength), at the same time decoupling from the dust temperature.

On the other hand, recent observational evidence for the temperature structure in starless cores is somewhat controversial. \cite{sysdiff} and \cite{tafalla1} have found an isothermal structure at around 9\,K for a sample of starless cores, including the L1544 and L1498 cores. Similarly, \cite{vandertak} have found from H$_2$D$^+$ observations that the gas temperature is nearly constant or at least does not drop inwards significantly in L1544. High-resolution observations have led \cite{crapsietal} to the conclusion that the gas temperature does go down in the very center of the L1544 core. A similar thermal structure (a few~K in the center and more than 10\,K in the envelope) is inferred for the L183 starless core by \cite{pagani} and for B68 globule by \cite{pinedabensch}. Quite surprisingly, a reversed temperature radial profile for B68 is found by \cite{b68reversed}. These authors argue that the envelope of this globule is actually colder than its central region.

These deviations from isothermality would have only a subtle effect on results of dynamical and chemical modeling. However, there is an aspect, which is more sensitive to temperature variations. In order to interpret observational data in terms of abundances and velocities, radiation transfer in molecular lines is commonly utilized. The synthetic line profiles depend critically on the radial distribution of the excitation temperature, which is a non-linear function of the kinetic temperature $T$. The importance of adequate temperature understanding has already been recognized, e.~g., by \cite{dfra_B68}. These authors found that utilization of the non-uniform temperature profile with decreasing $T$ at smaller radii increases abundances of C$^{18}$O, CS, and H$_2$CO, inferred for B68, by factors of 2--3 in comparison with the isothermal case.

In this Letter, we study the influence of a non-uniform $T$ distribution on synthetic integrated intensity $J_{\rm int}$ radial profiles for various CO and CS (and their isotopomers) transitions, using the L1544 and L1521E starless cores as templates. Observational data are taken from \cite{sysdiff} and \cite{tafl1521e}. The radiation transfer model as well as the assumed pre-stellar core chemical and thermal structure are briefly outlined in \S~2. Results for the detailed chemical modeling are presented in \S~3. Influence of the thermal structure on the inferred analytical abundance representations is discussed in \S~4.
Results are summarized in \S~5.

\section{Model}

Synthetic radial profiles of the integrated intensities in a prestellar core are computed with the Monte-Carlo code URAN(IA), described by \cite{pavshus}. We utilize density distributions for the L1544 and L1521E cores, represented in \cite{sysdiff} and \cite{tafl1521e} as $n(r)=n_0/(1+(r/r_0)^p)$, with $p=2.4$, $r_0=2800$~AU, $n_0=1.35\times10^6$~cm$^{-3}$ for L1544 and $p=2$, $r_0=4200$~AU, $n_0=2.7\times10^5$~cm$^{-3}$ for L1521E. A model core is assumed to be static, with a uniform radial microturbulent velocity $v_{\rm t}$ of 140 m~s$^{-1}$. We tried other $v_{\rm t}$ values as well and found that our conclusions do not depend on particular assumptions about this quantity.

\begin{figure}
\plotone{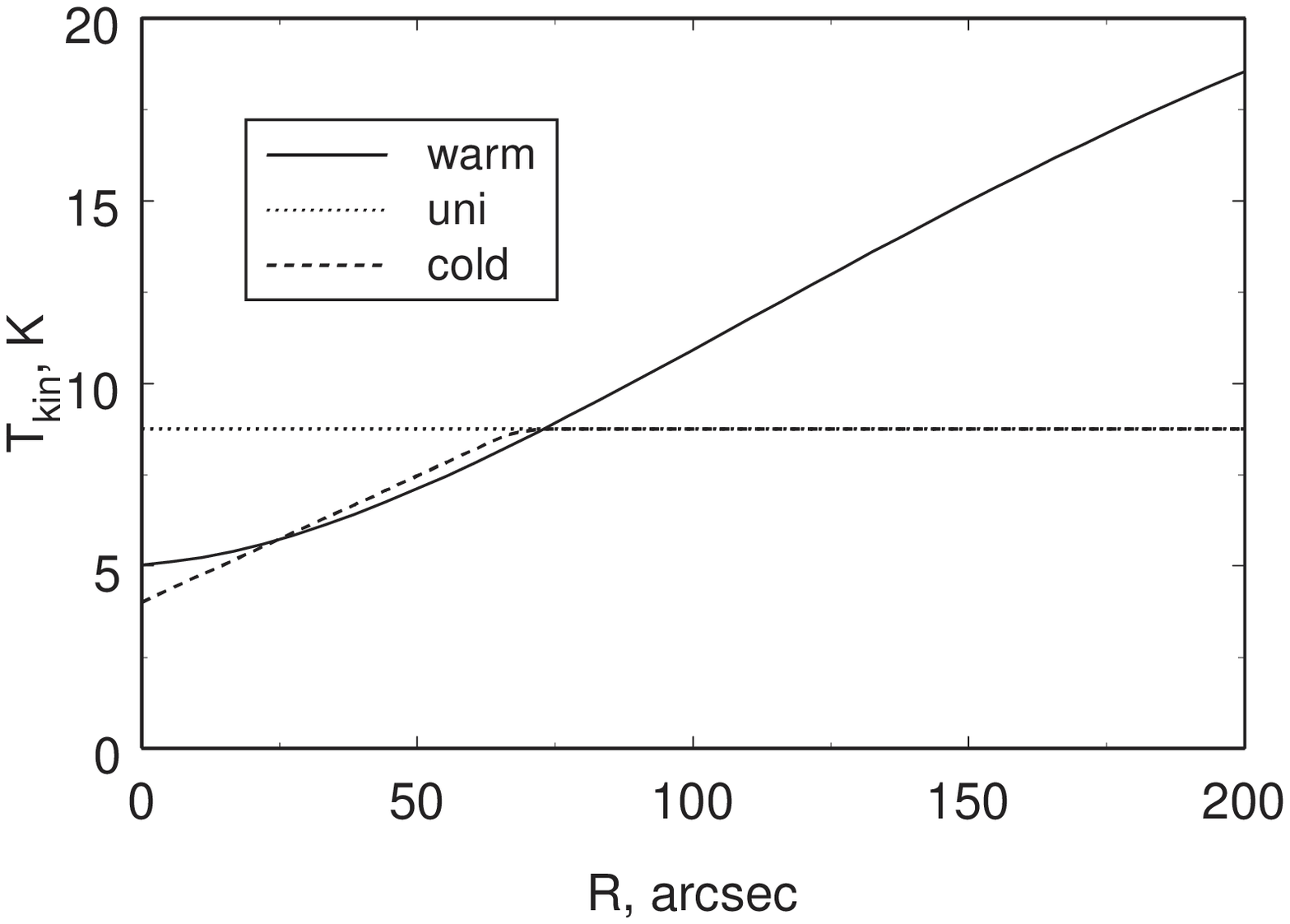}
\caption{Model temperature profiles. In the `uni' profile case, the temperature is uniform across the core at 8.75\,K for L1544 and 10\,K for L1521E (not shown). The `cold' profile corresponds to the cold center, while in the `warm' case, the envelope is warmer than in the `uni' case.}
\label{tprof}
\end{figure}

Three different temperature profiles are considered (Figure~\ref{tprof}). In models, designated `uni', $T$ is assumed to be 8.75\,K in L1544-based runs and 10\,K in L1521E-based runs at all radii. In `warm' models $T$ decreases from 15\,K at the core surface to about 5\,K at $r=0$ (the core center). The `cold' temperature profile is flat with $T=8.75$\,K (L1544) or 10\,K (L1521E) at $r/r_{\rm outer}>0.2$ and linearly decreases at smaller radii to 4\,K at the center.

Two different approaches for estimating the chemical structure of a starless core are employed. In \S~3 the CO and CS abundances are obtained through a detailed chemical modeling of a core. We use the same chemical model as in \cite{cb17} and the density profile for the L1544 core. The core is assumed to be isothermal ($T=10$\,K). We allow the chemistry to evolve for 2~Myr and then use the resulting molecular abundances at various times to model the emergent line profiles. Because of the assumption of uniform $T=10$\,K in the chemical models, they are formally inconsistent with the radiation transfer models. However, we calculated the same chemical models with the above non-uniform temperature profiles, and confirmed that the chemistry is not affected by these relatively small $T$ variations.

In \S~4, we use the density profiles for L1544 and L1521E and approximate the radial abundance profiles with pre-defined analytic expressions. Only C$^{18}$O(1--0) (L1544) and C$^{18}$O(2--1) (L1521E) transitions are considered in that section.

\section{Line intensities with complete chemistry}

\begin{figure}
\plotone{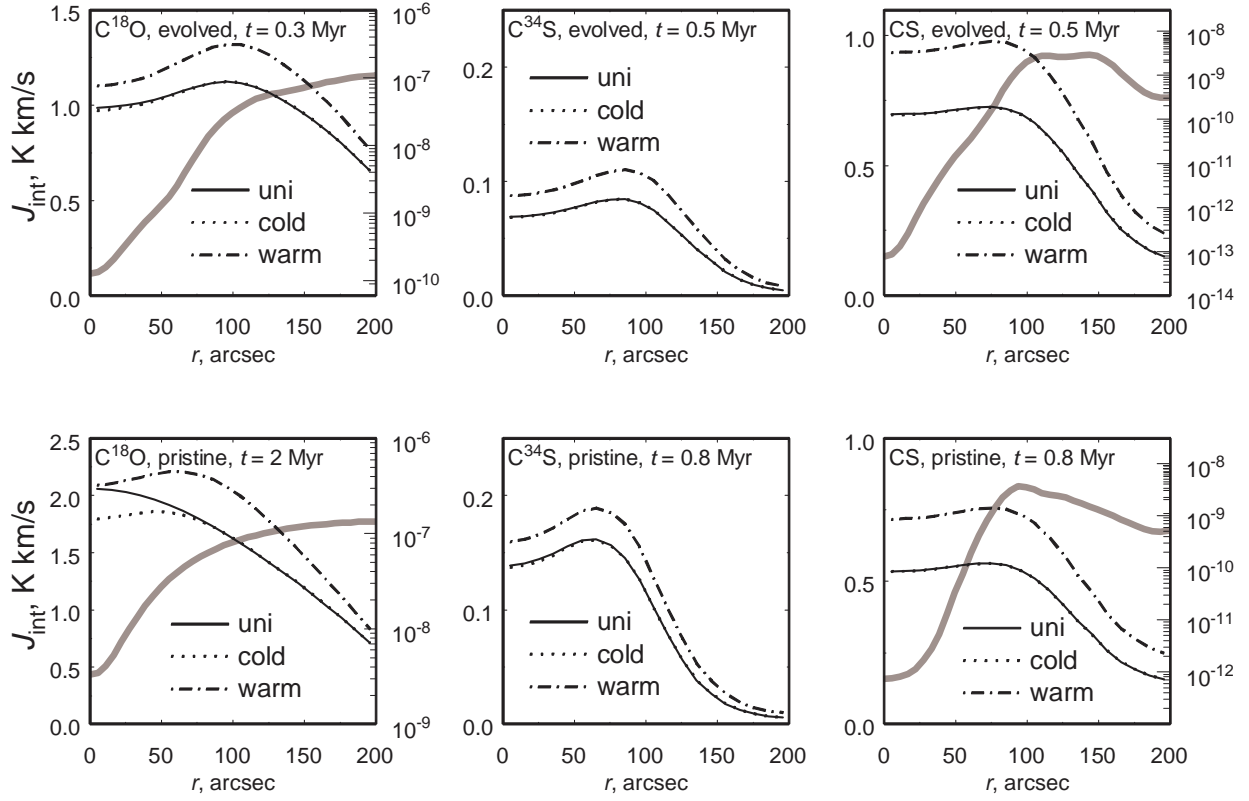}
\caption{Integrated intensity profiles for C$^{18}$O(1--0), C$^{34}$S(2--1), and CS(2--1) transitions, calculated with different assumptions on the core thermal structure. Thick grey lines and right axes represent the relative abundances of C$^{18}$O and CS. Shown are results for the high depletion case (top row) and the low depletion case (bottom row). Time moments are chosen, which provide the best agreement with L1544 observations in the `uni' model, except for the bottom left panel, where results at the end of the computation are used (see text).}
\label{jintsum}
\end{figure}

In the chemical modeling, we use two parameter sets. In the first set we assume a low sticking probability of $S=0.3$ (low depletion case). The model core is illuminated by interstellar UV field with $G=0.2$ and penetrated by cosmic rays with an ionization rate of $\zeta=3\times10^{-17}$~s$^{-1}$. This case is intended to represent ``chemically pristine'' starless cores like CB17 \citep{cb17} or L1521E \citep{hirota}. The other parameter set is more appropriate for ``chemically evolved'' cores (like L1544 or L1498). In this high depletion case, we adopt $S=1$, $\zeta=3\times10^{-18}$~s$^{-1}$ and the same UV field. Integrated intensity profiles for the considered transitions, temperatures, and chemical
parameters, convolved with the antenna beams, corresponding to observations in \cite{sysdiff}, are shown in Figure~\ref{jintsum}. The CO abundance in the low depletion case is not significantly time-dependent, so results after 2~Myr of evolution are shown. In the other plots, time steps are chosen, which in the `uni' model produce results close to observational data for L1544 \citep{sysdiff}.

In the high depletion case, all molecules are heavily depleted ($X({\rm max})/X(r=0)\approx200$ for CO), and it is of no surprise that the `uni' and `cold' models do not differ from each other. Both optically thin (C$^{18}$O, C$^{34}$S) and optically thick lines are generated at intermediate radii and are not affected by the temperature in the center. However, a warmer envelope (`warm' model) results in a brighter emission, with the effect being the strongest for the optically thick CS~(2--1) transition. This may have a direct impact on the analysis of observational data. If we use the adopted chemical model to interpret this emission erroneously as arising from an {\em isothermal\/} object, we would overestimate the core age by a factor of 1.5.

In the low depletion case with uniform $T$  (Figure~\ref{jintsum}, bottom left), the central hole in the C$^{18}$O integrated intensity distribution does not show up at all. Even though the depletion is still quite significant ($X({\rm max})/X(r=0)\approx60$), the absolute molecule number density peaks toward the center. Also, the uniform $T$ results in a relatively high excitation temperature. However, the central depression in $J_{\rm int}$ readily appears if we adopt a non-uniform temperature profile, with $T$ decreasing toward the center or increasing in the envelope. In the former case, the ring-like intensity profile arises because of the weaker central emission, in the latter case it is caused by the brighter envelope.

The temperature dependence of the C$^{34}$S and CS profiles is much weaker, because these profiles are mainly formed in the core envelope. As in the high depletion case, the isothermal model and the model with a cold center do not differ from each other, while the emission in the warmer envelope model is brighter than in other models.

Thus, whether or not the moderate depletion is visible as a central hole in the C$^{18}$O (1--0) $J_{\rm int}$ distribution, depends on the adopted thermal structure. A chemically evolved core always looks like a chemically evolved core, while a less evolved core may appear to be chemically mature just because its temperature structure is not taken into account properly.

\begin{figure}
\epsscale{0.6}
\plotone{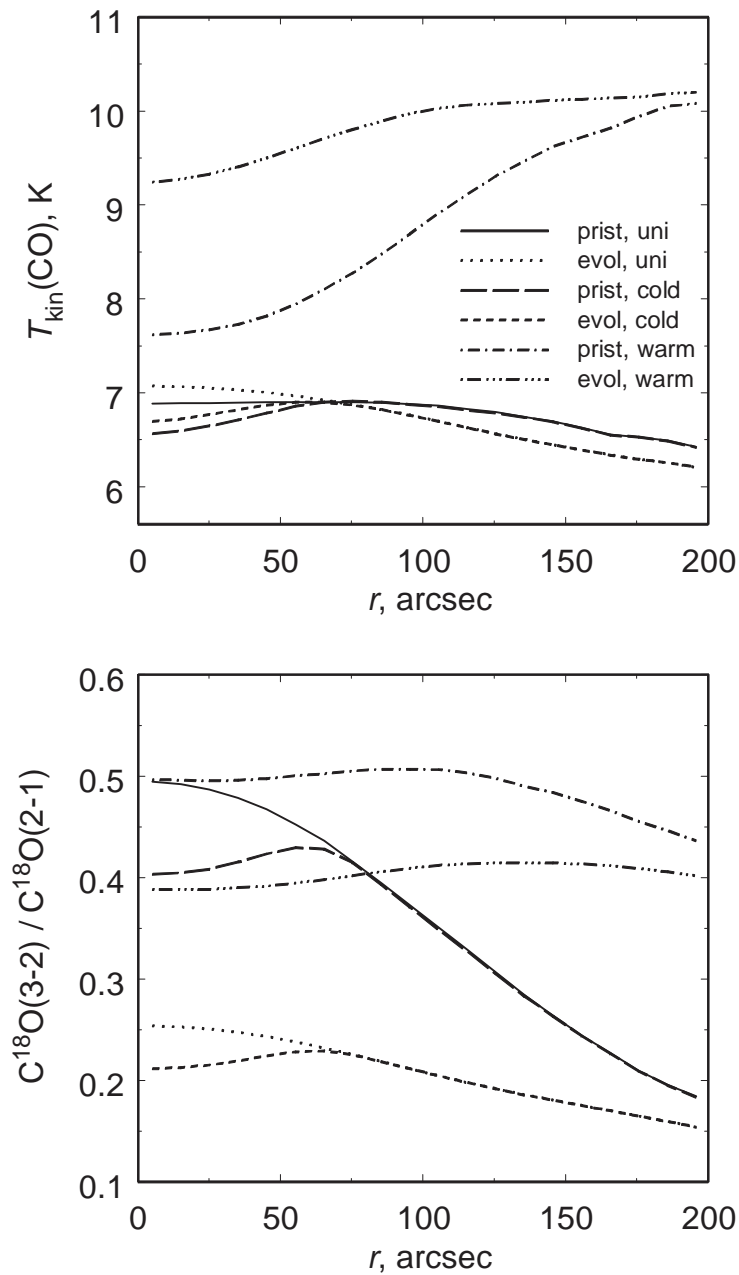}
\caption{Top: Temperatures computed from C$^{18}$O(2--1)/C$^{18}$O(1--0) integrated intensity ratios, assuming these transitions are thermalized and optically thin. Bottom: C$^{18}$O(3--2)/C$^{18}$O(2--1) intensity ratios. Abundances for $t=2$~Myrs are used.}
\label{exct}
\end{figure}

One may suggest that the temperature structure can be inferred from transition ratios, but in this case the interpretation is also not straightforward. In Figure~\ref{exct} (top panel) we show the $T$ profiles which would have been derived from the ratios of integrated intensities C$^{18}$O(2--1)/C$^{18}$O(1--0), assuming these transitions are thermalized and optically thin. Cores with warmer envelopes do stand out on this plot, but the C$^{18}$O(2--1)/(1--0) ratios do not allow to distinguish between `cold' and 'uni' models. In particular, the central temperature drop in `cold'  models is only formally present.

Some hints can, possibly, be extracted from the analysis of higher transitions. On the bottom panel of Figure~\ref{exct} we show the computed C$^{18}$O(3--2)/C$^{18}$O(2--1) integrated intensity ratios. In the pristine models the (2--1) line is saturated (i.e., its optical depth is greater than~1), while the (3--2) transition is (moderately) optically thin. Because of that, higher density at the core center enhances the (3--2) transition relative to the (2--1) transition. Thus, the C$^{18}$O(3--2)/C$^{18}$O(2--1) ratio is higher at small $r$ in `uni' and `cold' pristine models. In `warm' models the computed C$^{18}$O(3--2)/C$^{18}$O(2--1) ratio is large because in the warm envelope collisions more effectively populate higher levels.

In `uni' and `cold' evolved models , both C$^{18}$O(2--1) and C$^{18}$O(3--2) lines are optically thin, and the (2--1) transition does not show the saturation effects. At the same time, the (3--2) line is weak due to low density at the periphery and significant depletion at the core center. Thus, the C$^{18}$O(3--2)/C$^{18}$O(2--1) ratio is low.

Note the significant difference in C$^{18}$O(3--2)/C$^{18}$O(2--1) ratios between the `cold' and `uni' pristine models in the low depletion case (Figure~\ref{exct}, bottom panel, solid and long-dashed curves). This implies that in surveys of ``chemically young'' starless cores, which are currently underway, it is still desirable to pay specific attention to higher transitions in order to assess the thermal structure of these objects.

Another issue, which has been brought to our attention by the referee, is related to isotope selective photodissociation, which preferentially destroys C$^{18}$O. This process is not taken into account in our model, but it may affect the ability of C$^{18}$O to trace warm envelopes of starless cores. From this point of view, $^{13}$CO transitions may be better envelope probes.

\section{Analytical fits}

To study the sensitivity of derived molecular abundances on the temperature structure, we applied the technique, utilized by Tafalla and co-authors for L1544 \citep{sysdiff} and L1521E \citep{tafl1521e}, i.~e., we approximate the molecular abundance profiles with $X(r) = X_0 \exp\left[-n(r)/n_{\rm d}\right]$ and combine them with the three $T$ profiles described above.

{\em L1544}---L1544 is a well known example of a starless core with some species, like CO, CS, C$_2$S, being strongly depleted. Thus, according to the results of \S~3, we do not expect details of the thermal structure to have any significant effect on the interpretation of observational data.

\begin{figure}
\plotone{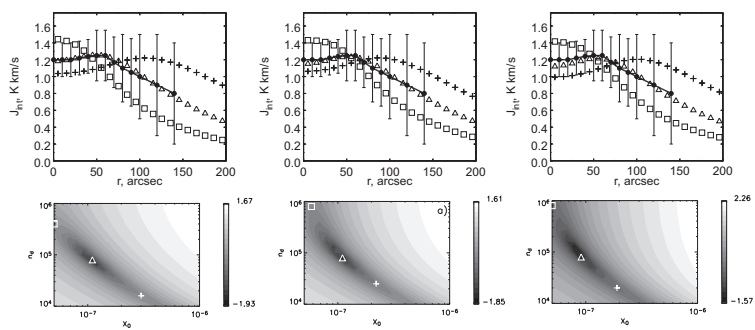}
\caption{Results of the analytical fits for `uni' (left column), `cold' (middle column), and `warm' (right column) models. Top: The observed (connected solid circles) and computed $J_{\rm int}$ profiles for the best-fit $X_0-n_{\rm d}$ combination (triangles) as well as for bracketing combinations marked with crosses and boxes on bottom panels. Bottom: $X_0-n_{\rm d}$ diagrams indicating the quality of agreement with observations. The darkest color corresponds to the best agreement. The best-fit positions for each $T$ profile are also marked with triangles.}
\label{diagr}
\end{figure}

With each of the three temperature profiles, we computed a series of RT models 
for different values of $X_0$ and $n_{\rm d}$ and compared the resultant distributions of the
integrated intensity with the observed distributions. In Figure~\ref{diagr} we present the
results of the comparison as a function of $X_0$ and $n_{\rm d}$. The best-fit positions in each
plot are indicated by triangles. Combinations that bracket the scatter in observational points for the C$^{18}$O(1--0) transition are shown with crosses and boxes. The error bars in Figure~\ref{diagr} represent the extent of scatter
in the observed integrated emission as a function of core radius \citep[e.g. Fig.~3 in][]{sysdiff}.

All the considered models do formally agree on the inferred values of $X_0\approx10^{-7}$ and $n_{\rm d}=8\times10^4$~cm$^{-3}$. However, the overall range of these parameters allowed by the scatter in $J_{\rm int}$ differs for various models. It appears that observations are marginally consistent with abundance profiles in which CO molecules are either totally depleted (crosses) or very moderately depleted (boxes). As the scatter mostly reflects the complex internal structure of the core, a difference between the bracketing $X_0-n_{\rm d}$ combinations may, to a certain degree, serve as a quantitative manifestation of a chemical inhomogeneity of the core.

{\em L1521E}---Unlike L1544, the starless core L1521E seems to be richer in gas phase CO, which is interpreted as a sign of its chemical youth \citep{hirota,tafl1521e}. Indeed, if we use the same exponential $X(r)$ profile as before and a uniform $T$ of 10\,K, we find the best agreement with observations of the C$^{18}$O(2--1) transition for $X_0=10^{-7}$ and $n_{\rm d}=10^6$~cm$^{-3}$. This $n_{\rm d}$ value is greater than the central density $n_0$ of L1521E, i.~e. the abundance distribution is nearly uniform. This is why in the `uni' model our conclusions are identical to those obtained by \cite{tafl1521e}.

\begin{figure}
\plotone{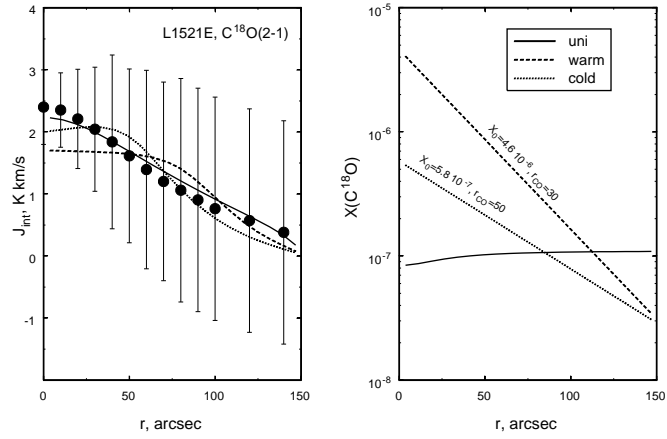}
\caption{Results of the analytical fits for the L1521E core. Radial abundance profiles which produce the best fit to the data are shown in the right panel with solid line (`uni' model), dashed line (`warm' model), and dotted line (`cold' model). Corresponding $J_{\rm int}$ profiles are shown in the left panel with the same line styles. Observational data are taken from \cite{tafl1521e}.}
\label{l1521e}
\end{figure}

However, the fit is different if we assume `cold' or `warm' $T$ profiles. Neither centrally depleted nor constant abundances are able to reproduce the observed $J_{\rm int}$ in L1521E. In order to check other possible inferences for the L1521E chemical composition, we employed a different $X(r)$ profile as well, assuming that $X(r)=X_0\exp(-r/r_{\rm CO})$. In Figure~\ref{l1521e} we compare the fitting results for the C$^{18}$O(2--1) transition in the L1521E core to the observational data, given in \cite{tafl1521e}. A combination of uniform $T$ and nearly constant abundance is able to fit the observations almost perfectly (solid line). To reproduce the observed $J_{\rm int}(r)$ in a core with a non-uniform $T$ distribution, one need a centrally peaked abundance profile (Figure~\ref{l1521e}). Again, like in a non-evolved chemical model, in models with lower central $T$ we may get either flat or even depressed $J_{\rm int}(r)$ despite the fact that the underlying abundance is growing toward the core center. The fit quality is not high, especially, in the `warm' model, but this problem can be solved by using an abundance profile with central flattening (like $\exp[-r^2/r_0^2]$) or by constructing an adequate chemical model.

\section{Conclusions}

The idea of this study is to draw attention to the problems related to the thermal structure
of prestellar cores. We demonstrate that
realistic radial temperature profiles can produce quite different line intensities, even when
the underlying chemical structure is the same. Effects of a non-uniform $T$ distribution
can mimic effects of chemical evolution and, thus, may hamper the interpretation of observations.
This is especially true for ``pristine'' starless cores and is less important
for chemically evolved cores. 
Finally, we would like to note that in this study we only consider transitions
of CO and CS molecules. While ammonia observations represent a more conventional
tool to measure the temperature in the interstellar medium, they are, probably,
prone to the same kind of uncertainty.

\acknowledgments

This study is supported by the DFG grant ``Molecular Cloud Chemistry'' HE 1935/21-1. DW acknowledges support from the RFFI Grant 07-02-01031 and the Russian Science Support Foundation grant.

\clearpage

\clearpage

\clearpage

\clearpage

\clearpage

\begin{thebibliography}{Bergin et al.(2006)}

\bibitem[Bergin et al.(2006)]{b68reversed} Bergin, E. A., Maret, S., van der Tak, F. F. S., Alves, J., Carmody, S. M., \& Lada, Ch. J. 2006, \apj, 645, 369

\bibitem[Crapsi et al.(2007)]{crapsietal} Crapsi, A., Caselli, P., Walmsley, M. C., Tafalla, M. 2007, \aap, 470, 221

\bibitem[Di Francesco et al.(2002)]{dfra_B68} Di Francesco, J., Hogerheijde, M. R., Welch, W. J., \& Bergin, E. A. 2002, \aj, 124, 2749

\bibitem[Galli et al.(2002)]{galli} Galli, D., Walmsley, M., \& Gon\c{c}alves, J. 2002, \aap, 394, 275

\bibitem[Goldsmith(2001)]{goldsmith} Goldsmith, P. F. 2001, \apj, 557, 736

\bibitem[Hirota et al.(2002)]{hirota} Hirota, T., Ito, T., Yamamoto, S. 2002, \apj, 565, 359

\bibitem[Lesaffre et al.(2005)]{lesaffre} Lesaffre, P., Belloche, A., Chi\` eze, J.-P., Andr\'e, P. 2005, \aap, 443, 961

\bibitem[Pagani et al.(2007)]{pagani} Pagani, L., Bacmann, A., Cabrit, S., Vastel, C. 2007, \aap, 467, 179

\bibitem[Pavlyuchenkov \& Shustov(2004)]{pavshus} Pavlyuchenkov, Ya. N., Shustov, B. M. 2004,
Astronomy Reports, 48, 315

\bibitem[Pavlyuchenkov et al.(2006)]{cb17} Pavlyuchenkov, Ya., Wiebe, D., Launhardt, R., Henning, Th. 2006, \apj, 645, 1212

\bibitem[Pineda \& Bensch(2007)]{pinedabensch} Pineda, J. L., Bensch, F. 2007, \aap, 470, 615

\bibitem[Tafalla et al.(2002)]{sysdiff} Tafalla, M., Myers, P. C., Caselli, P., Walmsley, C. M., Comito, C. 2002, \apj, 569, 815

\bibitem[Tafalla et al.(2004)]{tafalla1} Tafalla, M., Myers, P. C., Caselli, P., Walmsley, C. M. 2004, \aap, 416, 191

\bibitem[Tafalla \& Santiago(2004)]{tafl1521e} Tafalla, M., Santiago, J. 2004, \aap, 414, L53

\bibitem[van der Tak et al.(2005)]{vandertak} van der Tak, F. F. S., Caselli, P., \& Ceccarelli, C. 2005, \aap, 439, 195

\bibitem[Zucconi et al.(2001)]{zucconi} Zucconi, A., Walmsley, C. M., \& Galli, D. 2001, \aap, 376, 650

\end{thebibliography}
\end{document}